\documentclass[conference]{IEEEtran}
\ifCLASSINFOpdf
  \usepackage[pdftex]{graphicx}
  \DeclareGraphicsExtensions{.pdf,.jpeg,.png}
\else
  \usepackage[dvips]{graphicx}
  \DeclareGraphicsExtensions{.png}
\fi
\begin{document}
%
\title{An Approach for Assessing Clustering of Households by Electricity Usage}



%
\author{\IEEEauthorblockN{Ian Dent\IEEEauthorrefmark{1}\IEEEauthorrefmark{3},
Tony Craig\IEEEauthorrefmark{2},
Uwe Aickelin\IEEEauthorrefmark{1} and
Tom Rodden\IEEEauthorrefmark{3} }
\IEEEauthorblockA{\IEEEauthorrefmark{1}Intelligent Modelling and Analysis Group, School of Computer Science\\
University of Nottingham,
Nottingham, NG8 1BB, UK\\ Email: psxid@nottingham.ac.uk, Phone:+44 115 846 6568 }
\IEEEauthorblockA{\IEEEauthorrefmark{2}The James Hutton Institute, Craigiebuckler,
Aberdeen, AB15 8QH, UK}
\IEEEauthorblockA{\IEEEauthorrefmark{3}Horizon Digital Economy Research Institute, University of Nottingham}}


\maketitle

\begin{abstract}
How a household varies their regular usage of electricity is useful information for organisations to allow accurate targeting of behaviour modification initiatives with the aim of improving the overall efficiency of the electricity network.  The variability of regular activities in a household is one possible indication of that household's willingness to accept incentives to change their behaviour. 

An approach is presented for identifying a way of representing the variability of a household's behaviour and developing an efficient way of clustering the households, using these measures of variability, into a few, usable groupings.

To evaluate the effectiveness of the variability measures, a number of cluster validity indexes are explored with regard to how the indexes vary with the number of clusters, the number of attributes, and the quality of the attributes. The Cluster Dispersion Indicator (CDI) and the Davies-Boulden Indicator (DBI) are selected for future work developing various indicators of household behaviour variability.

The approach is tested using data from 180 UK households monitored for over a year at a sampling interval of 5 minutes. Data is taken from the evening peak electricity usage period of 4pm to 8pm.

\end{abstract}


%
\IEEEpeerreviewmaketitle

\section{Introduction}


The UK electricity industry, with the roll out of smart meters by 2020, will shortly have available a massively increased amount of data from domestic households and will be able to offer a much greater range of marketing campaigns to domestic users with the goal of changing their behaviour for the benefit of the overall electricity network.

In order to effectively target the marketing campaigns, it is necessary to cluster households into a manageable number of useful groupings so that each grouping can be addressed differently. Currently the utility companies use demographic data (family size, house size, location, etc.) as the basis for the clusters. The work presented makes use of the electricity meter data (solely) to explore whether useful clusters can be obtained based on the household's behaviour.

One aspect that will be explored is the variability in regular behaviours within the household (e.g. are mealtimes generally at the same time each day or is there large variability in the timing?) and various attributes are defined from the electricity meter data that represent the variability in a useful way. Adding an attribute that accurately reflects an interesting behaviour will lead to better clustering.

To assess the differing attributes objectively, it is necessary to define a process for this assessment. As inputs to the clustering process are changed (for example, the attributes being used), a measure is needed to assess whether the resulting clusters are "better" in some sense.

Marketing experience provides some guidelines as to what clusters are useful in real life and successful identification of useful clusters will mean that the cluster partitions meet the criteria as suggested by \cite{sarstedt2011concise} based on work by \cite{dibb1999criteria} \cite{tonks2009validity} and \cite{Kotler2008}. This suggests that clusters should be:

\begin{itemize}
\item Substantial: The segments are large and profitable enough to serve.
\item Accessible: The segments can be effectively reached and served, which requires
them to be characterized by means of observable variables.
\item Differentiable: The segments can be distinguished conceptually and respond
differently to different marketing-mix elements and programs.
\item Actionable: Effective programs can be formulated to attract and serve the
segments.
\item Stable: Only segments that are stable over time can provide the necessary
grounds for a successful marketing strategy.
\item Parsimonious: To be managerially meaningful, only a small set of substantial
clusters should be identified.
\item Familiar: To ensure management acceptance, the segment's composition should
be comprehensible.
\item Relevant: Segments should be relevant in respect of the company's competencies
and objectives.
\item Compactness: Segments exhibit a high degree of within-segment homogeneity
and between-segment heterogeneity.
\item Compatibility: Segmentation results meet other managerial functions' requirements.
\end{itemize}

The initial work presented focuses on the compactness and differentiable criteria with the goal of the clusters found being actionable.

This work investigates a set of cluster validity indexes to be used to evaluate future work in deriving attributes that represent flexibility of behaviour and that could be used as the basis for clustering into useful, meaningful groupings.

\section{Background}
Various cluster validity indexes have been proposed for application to general clustering tasks. In addition, some particular measures have been proposed for evaluating the clustering of electricity load profiles such as Chicco et al \cite{chicco2002review} which  provides a summary of techniques for clustering load profiles into similar classes of customers. \cite{chicco2002review} also details a number of validity indexes that can be used to assess the quality of the clusters generated. 

\cite{tsekouras2008pattern} gives 6 different validity indexes by which generated clusters can be evaluated when assigning electricity load profiles to clusters. 

Previous work often focuses on using each validity index to determine the appropriate number of clusters but, in this study, will be used to measure differences between various collections of attributes describing the households in order to determine which provides the best clustering. As this is not how the measures are generally used, investigation is undertaken as to how each behaves as the attributes (kind and number) change in order that suitable indexes can be selected for evaluation.

The work defines a set of features (initially very few) in order to describe the flexibility of a household and uses these features as the basis for clustering. When generally matching load profiles the order of the features has some information for matching, but is not often used. For example, readings for 10:30am may follow those for 10:00am and information can be derived from how the electricity usage changes between the readings. However, this is not applicable when considering measures of flexibility within the household as an unordered set of attributes does not allow for meaningful calculation of differences.

Validity indexes can be grouped into 3 major categories; internal, external and relative
\cite{jain1988algorithms}. Other literature suggests just two kinds of indexes; internal and external. The classification of some of the indexes is debatable, particularly regarding the difference between relative and internal indexes. One differentiating factor is the need for internal and external indexes to rely on statistical testing whereas relative indexes do not require the statistical testing and, hence, can be more computationally efficient.

External indexes are those used to compare the generated clusters with previously known information that has not been included in the clustering exercise. An example could be the clusters created from the NESEMP project using the demographic and attitudinal information collected. This information has not been included in the data mining clustering exercise but the clusters generated by the data mining can be compared to the demographic clusters using external indexes as a validation exercise.

Internal indexes are concerned solely with the internal representation of the generated clusters. An example may be an index calculated from the "tightness" of the members of a given cluster.

Relative indexes are intended to provide comparisons between different clustering solutions built using different input parameters. For example, relative indexes can be used to assess the "correct" number of clusters within a set of data analysed using kmeans. Each different k is used to generate a solution and then the relative index is used to decide between the differing solutions.

Each validity index has different ways of being used. Some are intended to be used so that the maximum value or the minimum value of the index is seen as the "best" solution whereas others need to be examined as to how the index value changes as the underlying input changes (e.g. the number of clusters). In the latter case, it is necessary to consider the differences between the index values as the variable being considered changes (or the differences between the differences).

\section{Approach}
\begin{figure}[h]
\centering
\includegraphics[width=3.4in]{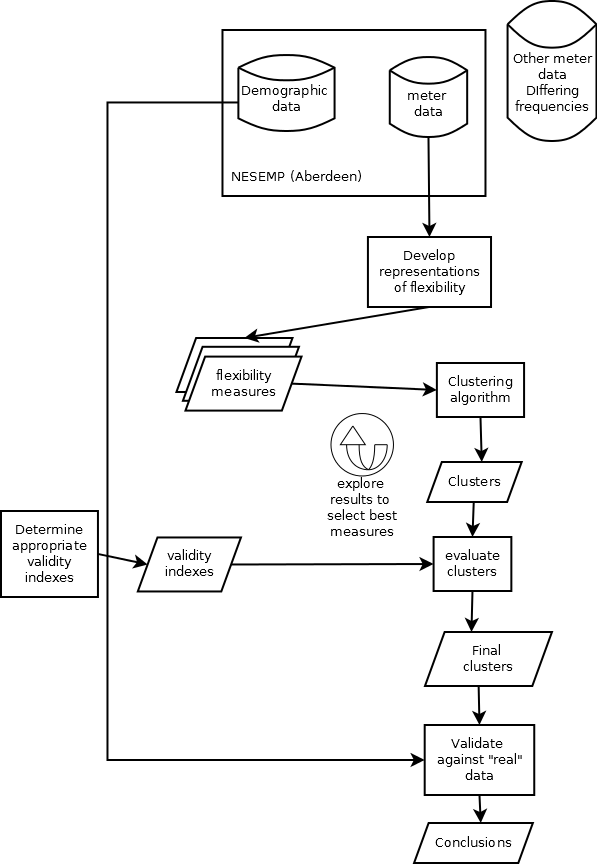}
\caption{Flow of analysis}
\label{analysis}
\end{figure}

A diagrammatic representation of the approach adopted can be seen at Figure \ref{analysis}.

The ongoing North East Scotland energy monitoring project (NESEMP) is examining the relationship between different types of energy feedback and psycho-social measures including individual environmental attitudes, household characteristics, and everyday behaviours.  As part of this project, several hundred households were monitored and the electricity usage was recorded every five minutes using CurrentCost monitors over a period of a year \cite{Craig2012}.

Data is used from 180 households within the NESEMP project providing about 17 million individual meter readings. The data is separated into that reflecting demographic and attitudinal aspects of the household's inhabitants and that collected directly from the electricity meter using the CurrentCost monitor installed in each house. The analysis concentrates solely on using the electricity meter with the demographic data being used in the final stage as the basis for a validation exercise.

Further data is available collected at various different sampling periods (from a few seconds to hourly) and, although not used in this exercise, will be used in future work to assess the sampling period necessary to derive useful information.

A subset of the data is selected based on the daily period of 4pm to 8pm and on days designated as "working days". This period represents the peak electricity usage period in the UK and it is likely to be the time when behaviour changes have the most useful effect on the overall electricity network.

Various measures can be derived from the electricity meter data and the goal of the work is to determine the most derived measures to use to usefully cluster the households. The focus is on representing flexibility of behaviour in the best way to allow interventions to be directed at households based on their measure of "flexibility". 

The data input to the clustering algorithms is of the format of a label (such as household+date) and a series of meter readings for times throughout the day. This data is summarised to provide a representative record for the given household (e.g. averaged over all weekdays) and then additional statistics derived from the meter data are added to the representative record (e.g. average time of peak usage).

This augmented representative record is then input to the clustering algorithm in order to derive useful clusterings of similar households. All the data in the record used for clustering is numeric - either meter readings or statistics derived from the readings. Some of the attributes in the record may be omitted from the clustering in some experiments. In particular, the averaged timed readings (e.g. the average reading at 4:05pm) are omitted from the following work in order to assess the applicability of the various cluster validity indexes when considering only the derived flexibility measures.

Flexibility measures are calculated for each household to give a statistic for the amount of variability during the evening period that the household displays with regard to the time of their maximum and their minimum usage. For example, for a given household, the time of maximum usage, represented by minutes after 4pm, during the peak evening period is found for each day of the sample. These times are then used to calculate the standard deviation of the times to give a measure of flexibility.

A measure of the total amount of energy used during the evening period for each household, averaged over all the days, is also calculated.

Data records input to the clustering algorithm consist of:
\begin{itemize}
\item household identifier
\item total usage measure
\item flexibility measure for time of maximum usage
\item flexibility measure for time of minimum usage
\end{itemize}

These records are normalised so that all values fit within 0 and 1 and are then input to the kmeans clustering algorithm to find 4 clusters.

Assessing the quality of the clusters requires the selection of appropriate cluster validity indexes which can be applied to the calculated clusters in order to reach a measure of how "good" the cluster is. Various validity indexes, and their behaviour as the underlying data changes, are considered below with a view to selecting the most appropriate for future work.
The definitions of the indexes are based on the following situation:

The data to be clustered consists of M records numbered as m=1,..M. Each record has H attributes numbered as h=1,..H.

The data is clustered into K clusters (numbered as k=1,..,K). Each cluster has $R_{k}$ members where $r_{(k)}$ is the rth record assigned to cluster k and $C_{(k)}$ is the calculated centre of the cluster k. $r_{(k)_h}$ is the hth attribute of record $r_{(k)}$.

The distance (d) between 2 profiles is defined as:
\begin{equation}
d(m_{i},m_{j}) = \sqrt{ \frac{1}{H} \sum\limits_{h=1}^H (m_i(h) - m_j(h))^2}
\end{equation}
where $m_i(h)$ and $m_j(h)$ are the hth attributes for two records, $m_i$ and $m_j$.

The infra-set distance $\hat{d}(S)$ of the members of a set, S with N members ($s_j$ where j=1,..,N) is defined as:
\begin{equation}
\hat{d}(S) = \sqrt{ \frac{1}{2N} \sum\limits_{n=1}^{N} \sum\limits_{p=1}^{N} d^{2}(s_n, s_p)}
\end{equation}

The scatter of a set of members of a cluster, SCAT(C), is defined as
\begin{equation}
SCAT(C_k) = \sqrt{ \frac{1}{R_k} \sum\limits_{r=1}^{R_k} \sum\limits_{h=1}^{H}
 (r_{(k)_h}, C_{(k)_h})^2 }
\end{equation}

\subsection{MIA}

The MIA (mean index adequacy) \cite{chicco2003customer} gives a value which relies on the amount by which each cluster is compact - i.e. if the members in the cluster are close together the MIA is low. 

\begin{equation}
MIA = \sqrt{ \frac{1}{K} \sum\limits_{k=1}^K \sum\limits_r d^2 (r_{(k)}, C_{(k)})}
\label{equation-eqn2}
\end{equation}

\subsection{CDI}

The CDI (cluster dispersion indicator) depends on the distance between the members of the same cluster (as for the MIA) but also includes the distances between the representative load diagrams for each cluster. This therefore measures both the compactness of the clusters and the amount by which each cluster differs from the others. A lower value for CDI suggests a better clustering solution.

\begin{equation}
CDI = \frac{1}{\hat{d}(C)} \sqrt{ \frac{1}{K} \sum\limits_{k=1}^K \hat{d}^2(R_k)}
\label{equation-eqn3}
\end{equation}
where C is the set of cluster centres and $R_{k}$ is the set of members of the kth cluster.

\cite{chicco2003customer} defines the CDI and uses it to select which attributes to include within the clustering exercise. 

\subsection{Similarity Matrix indicator}

The SMI is calculated by generating a matrix (of size K x K) where each element ($\alpha_{ij}$) is calculated as
\begin{equation}
\alpha_{ij} = \frac{1}{1 - \frac{1}{ln[d(C_{i},C_{j}]}}
\label{equation-eqn4}
\end{equation}
This matrix gives values for how similar each cluster centre is to each of the other cluster centres. 
The SMI is defined as the maximum $\alpha_{ij}$ where i$>$j.

The SMI gives a statistic dependent on the spread of the cluster centres found by the clustering algorithm but does not include any measure of the compactness of the members of each cluster \cite{chicco2003application}.

\subsection{Davies-Boulden indicator}

The DBI provides a measure of the ratio of the within cluster scatter to the between cluster separation \cite{davies1979cluster}. 

This measure has limited meaning when the clustering solution produces clusters containing single members. For correct use of this measure it is necessary to restrict the clustering algorithms to always produce clusters with at least 2 members. This can be done by restricting the algorithms or by identifying the "single member" clusters, treating them as outliers and omitting them from the study.

\begin{equation}
DBI = \frac{1}{K} \sum\limits_{j=1}^K max_i \frac{SCAT(C_i) + SCAT(C_j)}{d(C_i, C_j)}
where i \neq j
\end{equation}

\subsection{Ball and Hall} 
\cite{ball1965isodata} provides an index calculated as the average distance of each member of a cluster from the cluster centre.

\begin{equation}
Ball =  \frac{1}{K} \sum\limits_{k=1}^K \sum\limits_r d^2 (r_{(k)}, C_{(k)})
\label{equation-ball}
\end{equation}

\section{Experiments}

Experiments with differing numbers of clusters (varying between 2 and 20) are run using the kmeans clustering algorithm and the effect on the different cluster validity indexes is collected.

Similarly, keeping the number of clusters set at 4, additional attributes are added (up to a total of 8) in order to assess the effect of additional attributes on the cluster validity indexes. For the initial work, the additional attributes consist of random numbers drawn from the uniform[0,1] distribution. Future work will consider other descriptive measures relating to the households, which may be correlated with the existing attributes.

Again keeping the number of clusters as 4, the differing quality of attributes is tested by using 3 attributes drawn from the 8 possible attributes created in the previous step. The 3 attributes vary from being the 3 "real" attributes as defined above to being 3 random attributes. The random attributes will provide less useful information than the real attributes and the cluster validity indexes should reflect this.

The Cluster Validity Indexes are calculated for the derived clustering plan using the clustIndex package and custom developed functions. The R clustIndex package provides 15 internal/relative validity indexes that can be used to assess the quality of generated clusters. \cite{dimitriadou2002examination} provides background information on each index and guidelines on how it should be used.

The R package NbClust \cite{charrad2012package} provides a front end to various other R packages that calculate cluster validity indexes but with an emphasis of selecting the appropriate number of clusters.

\section{Results}

\subsection{Clustering results}
180 households were clustered into 4 clusters using the attributes of:
\begin{itemize}
\item total usage during the evening period
\item variability of time of maximum usage during the evening (as measured by the standard deviation of the time of peak usage in minutes after 4pm)
\item variability of time of minimum usage during the evening
\end{itemize}

\begin{figure}[h]
\centering
\includegraphics[width=3.4in]{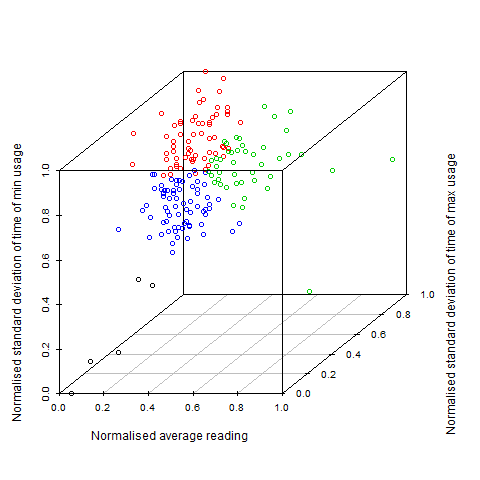}
\caption{4 clusters using kmeans with 3 attributes}
\label{3dclusters}
\end{figure}

The 4 clusters defined by kmeans can be seen at Figure \ref{3dclusters}. The results show 4 well defined clusters which can usefully be used for targeted marketing of interventions to change household behaviours. Some of the households appear as outliers (which may be due to some households having few days of data) and further work will identify additional preprocessing to remove unrepresentative data.

\begin{figure}[h]
\centering
\includegraphics[width=3.4in]{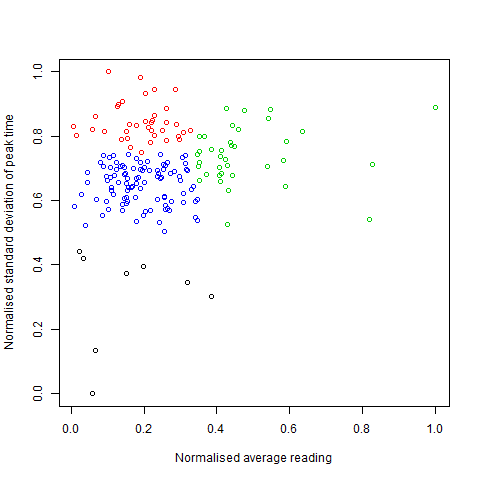}
\caption{4 clusters using kmeans with 2 attributes}
\label{2dclusters}
\end{figure}

The clustering is repeated using just 2 attributes (omitting the variability in time of minimum usage) and the results are shown at Figure \ref{2dclusters}. Again, 4 useful clusters can be seen and interventions for households with high usage during the peak period will be best directed at the households represented in green. Those interventions aimed at households with high variability in their behaviour are best directed at the households in red.

The households represented in black show little variability in behaviour and relatively low usage of electricity and it is likely they should be omitted from any initiatives. Those households represented in blue are those with lower total usage and a middling degree of variability and would probably be addressed as a lower priority after the red and green households.

These results give the base situation which is then modified in the following experiments by varying the inputs to the clustering algorithm.

\subsection{Number of clusters}

The cluster validity indexes calculated for differing numbers of clusters are summarised in Table \ref{table-measures}. Note that clustering partitions with single member clusters causes errors with the clustIndex package and this explains the lack of results for cluster numbers above 11 for DBI and Ball.

\begin{table}[ht]
\begin{center}
\caption{Validity measures for varying numbers of clusters}
\label{table-measures}
\begin{tabular}{rrrrrr}
  \hline
 clusters & mia & cdi & smi & dbi & ball \\ 
  \hline
2 & 1.18 & 77.29 & 0.64 & 0.31 & 4.18 \\ 
3 & 0.84 & 21.23 & 0.64 & 0.24 & 2.12 \\ 
4 & 0.64 & 6.65 & 0.64 & 0.19 & 1.24 \\ 
5 & 0.53 & 3.54 & 0.66 & 0.16 & 0.84 \\ 
6 & 0.44 & 2.08 & 0.67 & 0.15 & 0.58 \\ 
7 & 0.38 & 1.51 & 0.69 & 0.15 & 0.44 \\ 
8 & 0.34 & 1.09 & 0.69 & 0.15 & 0.35 \\ 
9 & 0.31 & 0.80 & 0.69 & 0.15 & 0.28 \\ 
10 & 0.28 & 0.63 & 0.71 & 0.14 & 0.23 \\ 
11 & 0.25 & 0.49 & 0.71 & 0.00 & 0.00 \\ 
12 & 0.23 & 0.39 & 0.71 & 0.00 & 0.00 \\ 
13 & 0.22 & 0.33 & 0.72 & 0.00 & 0.00 \\ 
14 & 0.20 & 0.27 & 0.71 & 0.00 & 0.00 \\ 
15 & 0.19 & 0.23 & 0.72 & 0.00 & 0.00 \\ 
16 & 0.18 & 0.20 & 0.72 & 0.00 & 0.00 \\ 
17 & 0.17 & 0.17 & 0.72 & 0.00 & 0.00 \\ 
18 & 0.16 & 0.15 & 0.72 & 0.00 & 0.00 \\ 
19 & 0.15 & 0.13 & 0.72 & 0.00 & 0.00 \\ 
20 & 0.14 & 0.12 & 0.72 & 0.00 & 0.00 \\ 
   \hline
\end{tabular}
\end{center}
\end{table}

It can be seen that most of the indexes change in the same way (i.e. either decreasing or increasing) as additional clusters are added. This means that a simple approach (e.g. picking the number of clusters with the index at maximum or minimum value) is not possible when searching for the "right" number of clusters, but examination of the changing size of the changes at each increase in numbers of clusters is required. This approach is similar to that of plotting the index against the number of clusters and then looking for an obvious "elbow" where there is a major change in order to suggest the appropriate number of clusters.

\subsection{Numbers of attributes}

\begin{table}[ht]
\begin{center}
\caption{Validity measures for varying numbers of attributes}
\label{table-measures-attr}
\begin{tabular}{rrrrrr}
  \hline
 number of attributes & mia & cdi & smi & dbi & ball \\ 
  \hline 
2 & 0.48 & 6.75 & 0.69 & 0.13 & 0.68 \\ 
3 & 0.64 & 6.65 & 0.64 & 0.19 & 1.24 \\ 
4 & 0.97 & 10.06 & 0.64 & 0.40 & 2.84 \\ 
5 & 1.29 & 11.31 & 0.58 & 0.44 & 4.96 \\ 
6 & 1.61 & 13.43 & 0.54 & 0.61 & 7.80 \\ 
7 & 1.93 & 15.58 & 0.52 & 0.82 & 11.21 \\ 
   \hline
\end{tabular}
\end{center}
\end{table}

When considering increasing the number of attributes, Table \ref{table-measures-attr} shows how the measures vary as attributes (consisting of random numbers) are added to the records input to the clustering algorithm with the number of attributes increasing from 2 up to 7.

The formulae for calculating the validity indexes are affected by the number of attributes used during the clustering and, as it is intended to assess the benefit of adding extra attributes, a measure that corrects for the number of attributes is needed. To investigate this the attributes in Table \ref{table-measures-attr} are divided by the number of attributes to produce the results in Table \ref{table-measures-attr-divide}.

\begin{table}[ht]
\begin{center}
\caption{Validity measures for varying numbers of attributes adjusted for number of attributes}
\label{table-measures-attr-divide}
\begin{tabular}{rrrrrr}
  \hline
 number of attributes & mia & cdi & smi & dbi & ball \\ 
  \hline 
  2 & 0.24 & 3.38 & 0.34 & 0.07 & 0.34 \\ 
  3 & 0.21 & 2.22 & 0.21 & 0.06 & 0.41 \\ 
  4 & 0.24 & 2.52 & 0.16 & 0.10 & 0.71 \\ 
  5 & 0.26 & 2.26 & 0.12 & 0.09 & 0.99 \\ 
  6 & 0.27 & 2.24 & 0.09 & 0.10 & 1.30 \\ 
  7 & 0.28 & 2.23 & 0.07 & 0.12 & 1.60 \\   
   \hline
\end{tabular}
\end{center}
\end{table}

The CDI validity index shows the most interesting behaviour as the only one that doesn't increase or decrease for each value for the number of attributes. With the adjustment for numbers of attributes the DBI and MIA are also of interest.

\subsection{Changing attributes}
In order to assess the effect of changing the quality of an attribute on the value of the validity indexes, a test was undertaken using the 3 attributes as above (total usage and variability of minimum and maximum times) and comparing this with using 3 random attributes created by sampling from the uniform distribution over 0:1. Four experiments were run with an additional random attribute replacing a real attribute at each experiment. The results can be seen in Table \ref{table-measures-random}.

\begin{table}[ht]
\begin{center}
\caption{Validity measures for varying randomness of attributes}
\label{table-measures-random} 
\begin{tabular}{rrrrrr}
  \hline
 random attributes & mia & cdi & smi & dbi & ball \\ 
  \hline
0 & 0.64 & 6.65 & 0.64 & 0.19 & 1.24 \\ 
1 & 0.81 & 9.49 & 0.62 & 0.29 & 1.96 \\ 
2 & 0.97 & 8.37 & 0.58 & 0.26 & 2.84 \\ 
3 & 1.26 & 10.27 & 0.54 & 0.37 & 4.75 \\ 
   \hline
\end{tabular}
\end{center}
\end{table}

All the indexes demonstrate worse values (either increasing or decreasing) as the quality of the attributes input to the clustering are made worse and a combination of the indexes is likely to be the best approach for testing different combinations of the same number of attributes.

\section{Conclusions}

In the absence of the demographic information, a relative index (or a combination of a number of relative indexes) should be used to compare differing input attributes in order to assess the attributes creating the "best" clusters. The demographic information will be introduced at the final stage of the analysis as a method of validating the derived clusters. 

No single validity index offers an obvious simple measure of the quality of a clustering partition scheme and, at best, only offers an indication as to how the quality of the cluster changes as the data or clustering algorithm parameters change. Most of the indexes investigated are intended for use in addressing the "how many clusters" question. For the intended work on finding flexibility measures, it is necessary to determine a cluster validity index (or a combination of a number) that allows the data input to the algorithm to be changed and for a measure of "quality" to be produced.

The CDI measure has been used in previous work \cite{chicco2003customer} to measure the quality of clusters as the attributes used in the clustering are changed and offers the most useful way forward. In addition, a simple combination of some of the other indexes (e.g. an average) may offer useful information particularly when comparing experiments with the same number of attributes. When the number of attributes is fixed and other input parameters (e.g. particular attributes) are used for clustering, then the Davies-Boulden index and the CDI will be used.

When exploring if adding an additional attribute is worthwhile with regard to improving the quality of the clusters, the CDI measure adjusted for number of attributes will be used.

\section{Future work}
Further pre-processing of the data will be done as the analysed data contains some households with relatively few data points and others that show up as outliers on the cluster graphs.

Additional data from other households is available and the clustering analysis will be repeated with this additional data allowing analysis of a total of 380 households from the NESEMP. Further datasets collected at differing frequencies (from 7 seconds to hourly) will also be considered to assess the effect of the sampling frequency on the quality of the clusters and to determine the minimum sampling frequency necessary to obtain useful clusters.

With the selection of the CDI and DBI validity indexes for evaluation of clustering schemes, differing attributes will be defined and used as the basis for clustering the households. Each set of attributes will be assessed using the validity indexes to produce an objective measure of whether particular attributes lead to better or worse cluster solutions. The attributes to add will include different representations of flexibility in regular behaviour as well as consideration of useful ratios such as night/day usage.

When considering changes to the input data such as varying the attributes that are included in the clustering, it is not possible to order the solutions (as it is, for example, when the number of clusters is increased which has an ordered sequence of possible numbers of clusters and where the differences in the cluster validity index can be assessed for adjacent values for the numbers of clusters). One approach for considering the various cluster validity indexes relating to differing combinations of attributes is to plot the changes and visually assess the "best" solution. Other solutions may be possible based on considering all possible combinations and assessing the benefit or otherwise of adding a particular attribute. 

As this initial work has been done with few attributes (3), the selection of appropriate validity indexes will be revisited once further possible attributes have been defined.


\section*{Acknowledgment}

This work is possible thanks to EPSRC grant references EP/I000496/1 and EP/G065802/1.

The work is part of a wider project to successfully apply demand side management techniques to gain benefits across the whole electricity network \cite{Kiprakis2011}.



\bibliographystyle{IEEEtran}
%
\bibliography{References}




\end{document}